\documentclass[11pt]{article}
\usepackage{epsfig}
\usepackage{amsfonts,latexsym}

\catcode`\@=11
\def\Let@{\relax\iffalse{\fi\let\\=\cr\iffalse}\fi}
\def\vspace@{\def\vspace##1{\crcr\noalign{\vskip##1\relax}}}
\def\multilimits@{\bgroup\vspace@\Let@
 \baselineskip\fontdimen10 \scriptfont\tw@
 \advance\baselineskip\fontdimen12 \scriptfont\tw@
 \lineskip\thr@@\fontdimen8 \scriptfont\thr@@
 \lineskiplimit\lineskip
 \vbox\bgroup\ialign\bgroup\hfil$\m@th\scriptstyle{##}$\hfil\crcr}
\def\Sb{_\multilimits@}
\def\endSb{\crcr\egroup\egroup\egroup}
\def\Sp{^\multilimits@}

\catcode`\@=11

\def\vereq#1#2{\lower3pt\vbox{\baselineskip1.5pt \lineskip1.5pt
\ialign{$\m@th#1\hfill##\hfil$\crcr#2\crcr\sim\crcr}}}

\newcommand{\be}[1]{\begin{equation}\label{#1}}
\newcommand{\ee}{\end{equation}}
\newcommand{\ba}[1]{\begin{eqnarray}\label{#1}}
\newcommand{\ea}{\end{eqnarray}}

\newcommand{\bmatrix}[1]{\left( \begin{array}{#1}}
\newcommand{\ematrix}{\end{array}\right)}

\newlength{\indentedwidth}
\newdimen\mathindent
\indentedwidth=\mathindent

\newenvironment{indented}{\begin{indented}}{\end{indented}}
\newenvironment{varindent}[1]{\begin{varindent}{#1}}{\end{varindent}}
\def\indented{\list{}{\itemsep=0\p@\labelsep=0\p@\itemindent=0\p@
   \labelwidth=0\p@\leftmargin=\mathindent\topsep=0\p@\partopsep=0\p@
   \parsep=0\p@\listparindent=15\p@}\footnotesize\rm}

\def\varindent#1{\setlength{\varind}{#1}%
   \list{}{\itemsep=0\p@\labelsep=0\p@\itemindent=0\p@
   \labelwidth=0\p@\leftmargin=\varind\topsep=0\p@\partopsep=0\p@
   \parsep=0\p@\listparindent=15\p@}\footnotesize\rm}

\begin{document}
\smallskip
\smallskip
\author{Alexey A. Kryukov \\
{University of Wisconsin}\\
\smallskip
{\small e-mail: aakrioukov@facstaff.wisc.edu}\\
\\
\smallskip
}

\title{Coordinate formalism on Hilbert manifolds}

\maketitle

\begin{abstract}
Infinite-dimensional manifolds modelled on arbitrary Hilbert spaces of functions
are considered. It is shown that changes in model rather than 
changes of charts within the same model make coordinate 
formalisms on finite and infinite-dimensional manifolds deeply similar. In this context the 
obtained infinite-dimensional counterparts of simple notions  
such as basis, dual basis, orthogonal basis, etc. are shown to be closely related to 
the choice of a model. 
It is also shown that in this formalism a single tensor equation on an infinite-dimensional
manifold produces a family of functional equations on different spaces of functions.
\end{abstract}

\bigskip

%

\section{Introduction}

\setcounter{equation}{0}

Methods of differential geometry have proven to be very useful in the theory
of gauge fields and especially in gravity. It is then natural to assume that
they must be even more important in the string theory, which aims to develop
into a ``Theory of Everything''.

Geometrization of string fields leads to infinite-dimensional manifolds.
These usually are manifolds of curves on a finite dimensional manifold. When
the strings are closed, they are the well known loop spaces.

Despite a very intense investigation of infinite-dimensional manifolds and many
results, up until now we do not understand clearly how to work
with them. In particular, we do not know what is an appropriate
generalization of the integration procedure on infinite-dimensional manifolds.
Probably related to this lack of knowledge is our trouble understanding
quantum theory based on the Feynman integral.

In addition we are not sure whether the manifolds that we consider a physically
acceptable. In particular, if we are serious about ``moving'' physics to an
infinite-dimensional background, the concept of ordinary space-time must be
derived from it. In this respect the manifold of curves on space-time is not
an appropriate concept as it needs the space-time to be defined in advance.

In such a situation any analysis bringing a new light on the notion of an
infinite-dimensional manifold is important.

In the paper a coordinate formalism on infinite-dimensional Hilbert
manifolds is developed and a coordinate version of tensor algebra is
constructed.

A Hilbert manifold is locally diffeomorphic to a separable Hilbert space. In
a most general scenario images of charts on an infinite-dimensional Hilbert
manifold belong to different Hilbert spaces. This fact is often neglected as
all such models are isomorphic. That is, we can assume that there is a fixed
Hilbert space model. The situation is similar in the case of a
finite-dimensional manifold. In the latter case images of charts belong to
n-dimensional vector spaces each isomorphic to $R^{n}$. There is, however,
an important difference. In applications we work with specific functional
representations of the elements of a Hilbert manifold. Functions behave very
differently with respect to any kind of analyzis. Therefore, a particular
choice of Hilbert model may become important.
As no particular Hilbert space is large enough to contain all functions
useful in applications, we are lead to consider several Hilbert models at
once.

From this there follows a more technical difference beween the finite and
infinite-dimensional settings. By fixing a chart on a finite-dimensional
manifold, we describe points by columns of numbers. When fixing a chart on a
Hilbert manifold instead, we describe points by elements of a Hilbert space 
$H$. These elements usually have a functional nature. They can be
represented, in turn, by columns of numbers, their components in a basis on 
$H$. 

The coordinate formalism developed in the paper seems to indicate a deep
connection between finite and infinite-dimensional manifolds. It is shown that
the choice of a Hilbert model even more than the choice of a chart within a 
given model is similar to the choice of coordinates on a finite-dimensional 
manifold. 

Here is the plan of the paper. To begin with, we analyze the notion of a
string in the string theory and describe a natural structure on the set
of strings. This leads to the idea of a string manifold, which is
a Hilbert manifold with a Riemannian metric and with charts taking values in
arbitrary Hilbert spaces of functions. We remark here that despite their
name, string manifolds at this stage have very little to do with the string
theory.

In section 3 the case of a linear string manifold and its conjugate is
carefully analyzed. In particular, the notions of a string basis, orthogonal
string basis, dual string basis, and coordinate space are introduced.

In section 4 we study linear transformation of coordinates on a linear string
manifold. An interesting example of coordinate transformations is carefully
analyzed. In particular, a possible meaning of the square of the Dirac $\delta $%
-function is explained.

Section 5 deals with the notion of a generalized eigenvalue problem. It is
shown in particular that each eigenvalue problem on a string space
generates a family of eigenvalue problems on different coordinate
spaces.

Section 6 generalizes the results to the case of nonlinear functional
coordinate transformations on a linear string manifold. 

Consideration of arbitrary local diffeomorphisms makes it possible
to define a string manifold structure. This is done in section 7. 
There we also discuss tensor bundles
and representation of tensors in local coordinates.


\section{What is a string?}

\setcounter{equation}{0}


The set of physical strings in the string theory is considered as a
factor-set $\Omega M=\Sigma /D$ of the set $\Sigma$ of (smooth) functions on
a given interval with values in a given space-time by the group $D$ of
diffeomorphisms of the interval. In other words, a string is a smooth curve
in space-time. The string fields are then functionals on strings. The
elements of $\Sigma$ are called parametrised strings.

As a geometric object a string is an analogue of a point of a
finite-dimentional manifold. A parametrized string $x^{\alpha }(s)$ is a
functional analogue of the set of coordinates of a point in a given
coordinate system. A parametrized string will be therefore called a
coordinate function of the string, or, simply, a coordinate of the string.

Reparametrization is the following transformation of the string coordinate
(index $\alpha $ is ommited): $\widetilde{x}(s)=x(\widetilde{s}(s))=\int
\delta (\widetilde{s}(s)-t)x(t)dt$. Here $\widetilde{s}=\widetilde{s}(s)$ is
a new parameter and the integral is understood as the convolution $\ast $,
i.e. $\int \delta (\widetilde{s}(s)-t)x(t)dt=\delta (\widetilde{s}(s))\ast
x(s)$. The finite-dimensional analogue of reparametrization is obtained by
replacing the $\delta $-function by the Kroeneker $\delta $-symbol, the
parameters $s$, $\widetilde{s}$ by the natural indices $i$, $j=j(i)$, and
the convolution by the sum. This gives $\widetilde{x}^{i}=\sum_{l}\delta
_{l}^{j(i)}x^{l}$. Therefore, reparametrization reduces to a permutation of
coordinates of the point.

Clearly, permutation is a very special transformation of coordinates and a
much larger group of general coordinate transformations preserves points of
a finite-dimensional manifold.

As strings are analogues of points of a finite-dimensional manifold, it is
natural to consider them as objects invariant under a functional analogue of
the group of general coordinate transformations. This motivates the idea to
consider strings as points of an infinite-dimensional differentiable
manifold $\it {S}$. That is, $\it {S}$ is locally diffeomorphic to an
infinite-dimensional Banach space $\bf {S}$.

To be able to measure distances and angles on $\it {S}$ we will endove $\it {S}$ with a
Riemannian structure. For this $\it {S}$ will be assumed to be a Hilbert
manifold, i.e. $\bf {S}$ will be a Hilbert space. As discussed in
introduction, the Hilbert model will not be fixed. So the model space $%
\bf {S}$ will be an abstract Hilbert space that can be identified with
an arbitrary Hilbert space of functions within a given family. This will permit,
in particular, to describe strings by singular functions (distributions) rather
than by only the smooth ones. The manifold $\it{S}$
will be called a string manifold. Admissible coordinate transformations
on a string manifold are then diffeomorphisms of open sets in Hilbert spaces
of functions.


\section{String space and its conjugate}

\setcounter{equation}{0}

To begin implementing this program, consider the case of a linear string
manifold $\bf {S}$, which we shall call a string space.

{\it Definition}. A {\it string space} $\bf {S}$ is an abstract vector space
that is also a differentiable manifold linearly diffeomorphic to an 
infinite-dimensional separable Hilbert space. 
In this case we also say that $\bf {S}$ is a {\it Hilbertable} space.

The fact that $\bf {S}$ is an abstract vector space means that it consists of
the elements of an unspecified nature. In particular, elements of two such
spaces can not be distinguished.

As any two infinite-dimensional separable Hilbert spaces are isomorphic, it follows that any 
two string spaces are linearly diffeomorphic. Throughout the paper the word ``isomorphism" 
will be used for a linear diffeomorphism of spaces.

As a result, any two string spaces have isomorphic
structures and identical elements. That is, we can assume that there is only one copy
of $\bf {S}$.

To develop a coordinate formalism on $\bf {S}$ we need to be able to 
identify $\bf {S}$ with a Hilbert space of functions.

{\it Definition}.  A {\it Hilbert space of functions} is either a Hilbert space 
$H$, elements of which are equivalence classes of maps between two given
subsets of $R^{n}$ or the Hilbert space $H^{\ast }$ dual to $H$. Two
elements $f,g\in H$ are called {\it equivalent} if the norm of $f-g$ in $H$ is
zero.

Consider a linear map ${e}_{H}:H\longrightarrow \bf {S}$ from a Hilbert space
of functions $H$ into the string space $\bf {S}$. The action of ${e}_{H}$
on $\varphi \in H$ will be written in one of the following ways: 

\begin{equation}
{e}_{H}(\varphi )=({e}_{H},\varphi )=\int {e}_{H}(k)\varphi (k)dk={e}%
_{Hk}\varphi ^{k}.
\end{equation}
The integral sign is used as a notation for the action of ${e}_{H}$ on an
element of $H$ and in general does not refer to an actual integration. We also
use an obvious and convenient generalization of the Einstein's summation
convention over the repeated indices $k$ one of which is above and one
below. Once again, only in special cases does this notation refer to an actual
summation or integration over $k$.

{\it Definition}.  A linear isomorphism ${e}_{H}$ from a Hilbert space $H$ 
of functions onto $\bf {S}$ 
will be called a {\it string basis} on $\bf {S}$.

It is clear that any string $\Phi $ is the image of a unique element $a\in H$,
i.e. 

\begin{equation}
{\Phi }={e}_{H}(a) 
\end{equation}
for a unique $a \in H$. Also, if 

\begin{equation}
{e}_{H}(a)=0, 
\end{equation}
then $a=0$. This justifies the definition of a basis. It is worth noticing
that the basis ${e}_{H}$ defines the space $H$ itself. In fact, it
acquires the meaning only as a map on $H$. On the other hand, within $H$ the
map ${e}_{H}$ can be choosen in different ways. 

Given ${e}_{H}$ the function $\varphi \in H$ such that $\Phi={e}_{H}\varphi$
will be called a {\it coordinate} (or an {\it $H$-coordinate}) of a string $\Phi \in {\bf {S}}$.
The space $H$ itself will be called a {\it coordinate space}.

Let $\pi _{H}:{\bf {S}}\longrightarrow H$ be a global linear coordinate
chart on $\bf {S}$ (which exists as $\bf {S}$ is isomorphic to a
Hilbert space). Then an obvious example of an $H$-basis is the linear
isomorphism $e_{H}=\pi _{H}^{-1}$. It is natural to call this
basis a {\it coordinate string basis} on $\bf {S}$.

Let ${\bf {S}}^{\ast}$ be the dual string space. That is, ${\bf {S}}^{\ast}$ 
is the space of all linear continuous functionals on strings. ${\bf {S}}^{\ast}$ 
can be considered as a string space with the chart 
$({\bf {S}}^{\ast },\pi _{H^{\ast }})$. Here $H^{\ast }$ is dual
of $H$ and $\pi _{H^{\ast }}$ is a linear isomorpism ${\bf {S}}^{\ast}$
onto $H^{\ast }$.

{\it Definition}.  A linear isomorphism of $H^{\ast }$ onto ${\bf {S}}^{\ast}$ 
will be called a {\it string basis on} ${\bf {S}}^{\ast}$.

We will denote such a basis by ${e}_{H^{\ast }}$. Decomposition of an
element ${F}\in \bf{{S}^{\ast }}$ with respect to the basis will be
written in one of the following ways:

\begin{equation}
F={e}_{H^{\ast }}(f)=({e}_{H^{\ast }},f)=\int {e}_{H^{\ast }}(k)f(k)dk=
{e}_{H^{\ast }}^{k}f_{k}.
\end{equation}

{\it Definition.} The basis ${e}_{H^{\ast }}$ will be called {\it dual} to the
basis ${e}_{H}$ if for any string ${\Phi }={e}_{Hk}\varphi ^{k}$ and for any
functional ${F}={e}_{H^{\ast }}^{k}f_{k}$ the following is true: 

\begin{equation}
{F}({\Phi })=f(\varphi ).
\end{equation}
In general case we have 

\begin{equation}
F(\Phi )=e_{H^{\ast }}f(e_{H}\varphi )=e_{H}^{\ast }e_{H^{\ast }}f(\varphi ),
\end{equation}
where $e_{H}^{\ast }:{\bf {S}}^{\ast }\longrightarrow H^{\ast }$ is the
adjoint of $e_{H}$. Therefore, $e_{H^{\ast }}$ is the dual string basis if $%
e_{H}^{\ast }e_{H^{\ast }}:H^{\ast }\longrightarrow H^{\ast }$ is the
identity operator. In this case we will also write

\begin{equation}
{e_{Hl}}^{\ast}e_{H^{\ast }}^{k}=\delta _{l}^{k}.
\end{equation}
In special cases $\delta _{l}^{k}$ is the usual Kroeneker symbol or the $%
\delta $- function.

The action of $F$ on $\Phi $ in any bases $e_{H}$ on $\bf {S}$ and $e_{H^{\ast}}$ 
on ${\bf {S}}^{\ast}$ will be sometimes written in the following way: 

\begin{equation}
F(\Phi )=e_{H^{\ast }}^{k}f_{k}e_{Hl}\varphi ^{l}=G(f,\varphi
)=g_{l}^{k}f_{k}\varphi ^{l},
\end{equation}
where $G$ is a non-degenerate bilinear functional on $H^{\ast }\times H$.

The dual basis always exists. In fact, the bilinear functional $G$ generates
a linear isomorphism $\widehat{G}:H^{\ast }\longrightarrow H^{\ast }$ by $(%
\widehat{G}f,\cdot )=G(f,\cdot )$. Therefore $\widehat{G}f=\widetilde{f}$
can be considered as a new $H^{\ast }$-coordinate of the dual string $F$.
This is the coordinate in the dual basis $\widetilde{{e}}_{H^{\ast }}$.

We have: 

\begin{equation}
F=(\widetilde{e}_{H^{\ast }},\widetilde{f})=(\widetilde{e}%
_{H^{\ast }},\widehat{G}f)=(\widehat{G}^{*}\widetilde{e}_{H^{\ast }},f)=({e}_{H^{\ast
}},f).
\end{equation}
Therefore

\begin{equation}
\widetilde{e}_{H^{\ast }}=\left( \widehat{G}^{*}\right) ^{-1}{e}_{H^{\ast }}.
\end{equation}
By definition the string space $\bf {S}$ is linearly isomorphic to a
separable Hilbert space $H$ of functions. Any linear isomorphism $\pi _{H}:{\bf {S}}\longrightarrow H$
induces the Hilbert structure on $\bf {S}$ itself. In fact, linear
structures on $\bf {S}$ and $H$ are the same. Also, let $\Phi ,\Psi \in 
\bf {S}$, $\varphi ,\psi \in H$ and $\varphi =\pi _{H}{\Phi , \psi} =\pi
_{H}{\Psi}$. Then define the inner product $(\cdot,\cdot)_{S}$ on $\bf {S}$ by 
$(\Phi ,\Psi )_{S}=(\varphi ,\psi )_{H}$, where $(\cdot,\cdot)_{H}$ is the inner
product on $H$. It is clear that with this inner product $\bf {S}$ is a
Hilbert space and $\pi_{H}$ becomes an isomorphism of Hilbert spaces. 
Respectively, whenever $\bf {S}$ is Hilbert we will assume that the string bases 
$e_{H}={\pi_{H}}^{-1}$ are isomorphisms of Hilbert spaces. That is,
a Hilbert structure on any coordinate space $H$ is induced by a choice of
string basis. In particular, two Hilbert spaces with the same elements can have
different inner products in which case they represent different coordinate spaces.

Let us assume that $H$ is a real Hilbert space. We have:
 
\begin{equation}
\label{orto}
(\Phi ,\Psi )_{S}={\bf{G}}(\Phi ,\Psi )=G(\varphi,\psi 
)=g_{kl}\varphi ^{k}\psi ^{l}. 
\end{equation}
Here $G:H\times H\longrightarrow R$ is a bilinear form defining the inner
product on $H$ and ${\bf {G}}:{\bf {S}}\times {\bf {S}}\longrightarrow R$
is the induced bilinear form. The expression on the right is a convenient
form of writing the action of $G$ on $H\times H$. 

{\it Theorem. }The choice of a coordinate Hilbert space determines the
corresponding string basis up to a unitary transformation.

{\it Proof. }Let $e_{H}$ and $\widetilde{e}_{H}$ be two string bases on 
$\bf {S}$ with the same coordinate space $H$. Then $(\Phi ,\Psi )_{S}=
G(\varphi ,\psi )=\widetilde{G}(\widetilde{\varphi },\widetilde{\psi })=
\widetilde{G}(U\varphi ,U\psi )$. As $\widetilde{G}=G$ we have $G(\varphi
,\psi )=G(U\varphi ,U\psi )$, that is, $U$ is a unitary transformation.
Therefore $e_{H}=\widetilde{e}_{H}U$, i.e. the basis $e_{H}$ is determined
up to a unitary transformation on $H$.

{\it Definition}.  A string basis ${e}_{H}$ in $\bf {S}$ will be called
{\it orthonormal} if 
\begin{equation}
(\Phi ,\Psi )_{S}=f_{\varphi}(\psi ),
\end{equation}
where $f_{\varphi}=(\varphi,\cdot)$ is a regular functional and $\Phi=e_{H}\varphi$,
$\Psi=e_{H}\psi$ as before. That is, 
\begin{equation}
(\Phi,\Psi)=f_{\varphi}(\psi)=\int \varphi(x)\psi(x)d\mu (x),
\end{equation}
where $\int$ here denotes an actual integral over a $\mu $-measurable set $D \in R^{n}$.

The bilinear form ${\bf{{G}:{S}}}\times {\bf {S}}\longrightarrow R$
generates a linear isomorphism $\widehat{\bf{G}}:{\bf {S}}
\longrightarrow {\bf{{S}}^{\ast }}$ by ${\bf{G}}(\Phi ,\Psi )=(\widehat{
\bf{G}}\Phi ,\Psi )$.
In any basis $e_{H}$ we have
\begin{equation}
(\Phi ,\Psi )_{S}=\widehat{\bf{G}}(e_{H}\varphi ,e_{H}\psi )=e_{H}^{\ast
}\widehat{\bf{G}}e_{H}\varphi (\psi ),
\end{equation}
where $e_{H}^{\ast }\widehat{\bf{G}}e_{H}$ maps $H$ onto $H^{\ast }$. 
If $e_{H}$ is orthonormal, then $e_{H}^{\ast }\widehat{\bf{G}}e_{H}\varphi
=f_{\varphi} $.
In this case we will also write 
\begin{equation}
\label{orto1}
(\Phi ,\Psi )_{S}=\varphi ^{k}\psi ^{k}=\delta _{kl}\varphi ^{k}\psi ^{l}.
\end{equation}
In a special case $\delta _{kl}$ can be the Kroeneker symbol or Dirac's
$\delta $-function $\delta (k-l)$.

It is important to realize that not every coordinate Hilbert space $H$ can
produce an orthonormal string basis ${e}_{H}$. Assume, for example, that $H$
contains the $\delta $-function as a coordinate $\varphi $ of a string $\Phi
\in \bf {S}$ (example of such $H$ is given below). Then $\delta
_{kl}\varphi ^{k}\varphi ^{l}$ is not defined, that is, $\delta $-function
is not a coordinate function of a string in orthonormal basis.

This does not contradict the well known existence of an orthonormal basis in
any separable Hilbert space. In fact, the meaning of a string basis is
quite different from the meaning of a classical basis on a Hilbert space.
Namely, a  string basis on $\bf {S}$ permits us to represent an invariant with
respect to functional transformations object (string) in terms of a
function, which is an element of a Hilbert space. A basis on a Hilbert space
in turn permits us to represent this function in terms of numbers, components
of the function in the basis.

Equation (\ref{orto1}) shows that orthonormality of a string basis imposes a 
symmetry between coordinates of the
dual objects in the basis. In particular, if $e_{H}$ is
orthonormal, then $H$ must be an $L_{2}$-space, i.e. a space $L_{2}(D,\mu)$
of square integrable functions on a $\mu$-measurable set $D \in R^{n}$. 
Thus, Hilbert spaces $l_{2}$ and $L_{2}(R)$ are examples of
coordinate spaces that admit an orthonormal string basis.

Notice that formula (\ref{orto1}) suggests that if $H$
possesses an orhonormal basis, then the chart $({\bf {S}},\pi_{H})$ is analoguous to
a rectangular Cartesian coordinate systems in Euclidean space. More 
general formula (\ref{orto}) shows that other coordinate Hilbert spaces 
produce analogues of oblique Cartesian coordinate systems in Euclidean space.

\section{Linear coordinate transformations on $\bf {S}$}

\setcounter{equation}{0}

{\it Definition.} A linear coordinate transformation on $\bf {S}$ is 
an isomorphism $\omega :\widetilde{H}\longrightarrow H$ of Hilbert 
spaces which defines a new string
basis $e_{\widetilde{H}}:\widetilde{H}\longrightarrow \bf {S}$ by 
$e_{\widetilde{H}}=e_{H}\circ \omega $. 

Let $\varphi $ be coordinate of a string $
\Phi $ in the basis $e_{H}$ and $\widetilde{\varphi }$ its coordinate in the
basis $e_{\widetilde{H}}$. Then $\Phi =e_{H}\varphi =e_{\widetilde{H}}
\widetilde{\varphi }=e_{H}{\omega} \widetilde{\varphi }$. That is, $\varphi
=\omega \widetilde{\varphi }$ by the uniqueness of the decomposition. This provides a
transformation law of string coordinates under a change of coordinates.

For the metric we have:

\begin{equation}
(\Phi ,\Psi )_{S}=(\widehat{G}\varphi ,\psi )=(\widehat{\widetilde{G}}\omega 
\widetilde{\varphi },\omega \widetilde{\psi }),
\end{equation}
where $\widehat{G}
:H\longrightarrow H^{\ast }$ and $\widehat{\widetilde{G}}:\widetilde{H}
\longrightarrow \widetilde{H}^{\ast }$ are operators defining inner products
on $H$ and $\widetilde{H}$. Then 

\begin{equation}
\widehat{G}=\omega ^{\ast }\widehat{\widetilde{G}}\omega ,
\end{equation}
where $\omega ^{\ast }$ is the adjoint of $\omega$.
This is the transformation law of the metric under a change of
coordinates.

By fixing $\widetilde{H}$ to be, say $L_{2}(R)$ we see that coordinate
representation of the metric in any coordinate space can be obtained from
the $L_{2}$-metric by means of isomorphism $\omega $.

Notice also, that $\widehat{G}^{-1}$ defines a metric on $H^{\ast }$. In
fact, if $f,g\in H^{\ast }$, $f=\widehat{G}\varphi $, $g=\widehat{G}\psi $,
we can define 

\begin{equation}
\label {dualmetric}
(f,g)_{H^{\ast }}=(\widehat{G}^{-1}f,g)=(\psi ,\varphi )_{H},
\end{equation}
which gives a metric on $H^{\ast }$.

{\it Example.} Let us see the string coordinate transformations in action.
Let $W$ be the Schwartz space of infinitely differentiable rapidly
decreasing functions on $R$. That is, functions $\varphi \in W$ satisfy
inequalities of the form $\left| x^{k}\varphi ^{(n)}(x)\right| \leq C_{kn}$
for some constants $C_{kn}$ and any $k,n=0,1,2,...$. Let $W^{\ast }$ be the
dual space of continuous linear functionals on $W$. Let us find a Hilbert
space which contains $W^{\ast }$ as a topological subspace. For this
consider a linear transformation $\rho $ from $W^{\ast }$ into $W$ given by 
$(\rho f)(x)=\int f(y)e^{-(x-y)^{2}-x^{2}}dy$ for any $f\in W^{\ast }$.

{\it Theorem.} $\rho (W^{\ast })\subset W$ and $\rho $ is injective.

{\it Proof.} It is known (see \cite{Gelfand1}) that every functional $f\in
W^{\ast }$ acts as follows:

\begin{equation}
(f,\varphi )=\int F(x)\varphi ^{(m)}(x)dx, 
\end{equation}
where $F$ is a continuous function on $R$ of power growth and $\varphi ^{(m)}(x)$ is 
the derivative of order $m$ of the function $\varphi(x) \in W$.
Therefore,

\begin{equation}
\int f(y)e^{-(x-y)^{2}-x^{2}}dy=\int F(y)\frac{d^{m}}{dy^{m}}%
e^{-(x-y)^{2}-x^{2}}dy.
\end{equation}
As $F$ is continuous of power growth, integration gives an
element of $W$.

To check injectivity of $\rho $ assume $\rho f=0$. Then

\begin{equation}
\int F(y)\frac{d^{m}}{dy^{m}}e^{-(x-y)^{2}}dy=0. 
\end{equation}
Differentiating an arbitrary number k of times under the integral sign and 
changing to $z=x-y$ we have:

\begin{equation}
\int F(x-z)\frac{d^{m+k}}{dz^{m+k}}e^{-z^{2}}dz=0. 
\end{equation}
Let us use the fact that

\begin{equation}
\frac{d^{m+k}}{dz^{m+k}}e^{-\frac{z^{2}}{2}}=H_{m+k}(z)(-1)^{m+k}e^{-\frac{z^{2}}{2}}, 
\end{equation}
where $H_{m+k}(z)$ are Hermite polynomials. It follows that
 
\begin{equation}
\int F(x-z)e^{-\frac{z^{2}}{2}}\varphi _{m+k}(z)dz=0, 
\end{equation}
where $\varphi _{m+k}(z)=H_{m+k}(z)e^{-\frac{z^{2}}{2}}$ is a complete
orthonormal system of functions in $L_{2}(R)$. Therefore, if $i>m$ all
Fourier coefficients $c_{i}(x)$ of the function $F(x-z)e^{-\frac{z^{2}}{2}
}\in L_{2}(R)$ are equal to zero. That is, almost everywhere, and by
continuity of $F$ everywhere, we have

\begin{equation}
F(x-z)e^{-\frac{z^{2}}{2}}=\sum_{i=0}^{m-1}c_{i}(x)H_{i}(z)e^{-\frac{z^{2}%
}{2}}. 
\end{equation}
Therefore $F$ is a polynomial function of $z$ of degree $m-1$. As $x-z=y$ it
follows that $F$ is a polynomial function of $y$ as well. Thus,

\begin{equation}
(f,\varphi )=\int P_{m-1}(y)\varphi ^{(m)}(y)dy, 
\end{equation}
where $P_{m-1}$ is a polynomial function of degree $m-1$. Integration by
parts $m$ times gives then $f=0$ proving injectivity of $\rho $.

Assume the strong topology on $W^{\ast }$. We could, however, choose the
weak topology as well as the weak and strong topologies on $W^{\ast }$ are
equivalent \cite{Gelfand1}. Let us define a topology on the linear space $%
\widetilde{W}$=$\rho (W^{\ast })$ by declaring open sets on $\widetilde{W}$
to be images of open sets on $W^{\ast }$ under the action of $\rho $. As $%
\rho $ is a bijection, this indeed defines a topology on $\widetilde{W}$. In
this topology $\rho $ and $\rho ^{-1}$ are continuous.

{\it Theorem.} The embedding $\widetilde{W}\subset W$ is continuous.

{\it Proof.} Topology on $W$ may be defined by the countable system of
norms

\begin{equation}
\left\| \varphi \right\| _{p}={\sup_{x \in R;k,q\leq p}}\left|
x^{k}\varphi ^{(q)}(x)\right| ,
\end{equation}
where $k,q,p$ are nonnegative integers and $\varphi ^{(q)}$ is the
derivative of $\varphi $ of order $q$. The strong topology on $W^{\ast }$ is
defined by taking as neighborhoods of zero the sets of functionals $f\in
W^{\ast }$ for which 

\begin{equation}
\label{strong}
{\sup_{\varphi \in B }}\left| (f,\varphi )\right|<{\epsilon}.
\end{equation}
Here $B\subset W$ is any bounded set (that is, a set bounded with respect to each norm $
\left\| {\cdot}\right\| _{p}$) and $\epsilon >0$ is any number.

To prove that the embedding $\widetilde{W}\subset W$ is continuous, we need
to show that for any neighborhood $O$\ of zero in $W$ a neighborhood 
$\widetilde{O}$ of zero in $\widetilde{W}$ can be found such that $\widetilde{
O}\subset O$. 

Let then $O$ be given by ${\sup_{x \in R;k,q\leq p}}\left| x^{k}\varphi
^{(q)}(x)\right| <K$ for some $p$ and $K$. \ Consider $O^{\ast } $ defined
by ${\sup_{\varphi \in B}}\left| (f,\varphi )\right| <{\epsilon}$.
Let us show that for some choice of $B$ and $\epsilon $ the neighborhood $%
\widetilde{O}=\rho (O^{\ast })$ is a subset of $O$.

For this, let 

\begin{equation}
\psi (x)=\int f(y)e^{-(x-y)^{2}-x^{2}}dy\in \widetilde{O}, 
\end{equation}
where $f\in O^{\ast }$, i.e. ${\sup_{\varphi \in B}}\left|
(f,\varphi )\right| <{\epsilon} $. Then 

\begin{equation}
\left\| \psi \right\| _{p}={\sup_{x\in R;k,q\leq p}}\left| x^{k}\psi
^{(q)}(x)\right| ={\sup_{x\in R;k,q\leq p}}\left| \int f(y)
x^{k}\left(e^{-(x-y)^{2}-x^{2}}\right) ^{(q)}dy\right| . 
\end{equation}
Consider $\varphi _{x}(y)=x^{k}\left(e^{-(x-y)^{2}-x^{2}}\right) ^{(q)}$. \
For every $x$ the function $\varphi _{x}(y)$ is in $W$. \ The set $A$ of
functions $\varphi _{x}$ parametrised by $x$ is bounded in $W$ with respect
to each norm $\left\| {\cdot}\right\| _{l}$ that define $W$. That is, $A$ is
bounded in $W$. Let then take $B=A$ and $\epsilon =K$. Then $\left\| \psi
\right\| _{p}<K$ and so $\psi \in O$. Therefore $\widetilde{O}\subset O$,
which proves the theorem.

{\it Theorem.} The embedding $W\subset W^{\ast }$ is continuous.

{\it Proof.} We assume here that every $\psi \in W$ is identified with
the regular functional $f_{\psi } \in W^{\ast}$ defined by $(f_{\psi },\varphi )=\int
\psi (x)\varphi (x)dx$.

If $O$ is a neighborhood of zero in $W^{\ast }$, then ${\sup_{\varphi \in B}}
\left| (f,\varphi )\right| <{\epsilon}$ for some bounded $B$ and some
$\epsilon $. Consider the neighborhood $\widetilde{O}\subset W$ of functions $%
\psi $ given by $\sup_{x}\left| \psi (x)\right| <K$. Then 

\begin{equation}
{\sup_{\varphi \in B}}\left| (f_{\psi },\varphi )\right| <K{\sup_{\varphi \in B}}
\int \left| \varphi (x)\right|dx =K\cdot L_{B} 
\end{equation}
for some constant $L_{B}$. Here we used the fact that any set bounded in $W$ 
is bounded in $L_{1}(R)$. Therefore taking $K=\frac{\epsilon }{L_{B}}$ we
obtain the desired inclusion $\widetilde{O}\subset O$.

Let us now narrow down the domain of the operator $\rho $ to the subspace $%
L_{2}^{\ast }(R)$ of $W^{\ast }$.

{\it Theorem.} The embedding $L_{2}^{\ast }(R)\subset W^{\ast }$ is
continuous.

{\it Proof.} Any neighborhood $O\subset W^{\ast }$
is given by ${\sup_{\varphi \in B}}\left| (f,\varphi )\right|
<\epsilon $ for some bounded set $B\subset W\subset L_{2}(R)$ and some 
${\epsilon} >0$. Also, since the embedding $W \subset L_{2}(R)$ is continuous,
any set bounded in $W$ is bounded in $L_{2}(R)$. 
Let then $A$ be a ball in $L_{2}(R)$ that contains $B$. Consider the 
neighborhood $\widetilde{O}$ in $L_{2}^{\ast }(R)$ given by 
${\sup_{\varphi \in A}}\left| (f,\varphi )\right| <{\epsilon}$. Then as 
$B \subset A$ we also have 
${\sup_{\varphi \in B}}\left| (f,\varphi )\right| <{\epsilon}$.  
We therefore obtain the desired inclusion $\widetilde{O}\subset O$.

In particular, the restriction $\rho |_{L_{2}^{\ast }}$ of $\rho $ on 
$L_{2}^{\ast }(R)$ is a continuous operator $\rho :L_{2}^{\ast
}(R)\longrightarrow W$ which we will denote again by $\rho $.

Let $H=\rho L_{2}^{\ast }(R)$. Then $\rho $ induces the Hilbert structure on 
$H$.

{\it Theorem.} With this Hilbert structure the embedding $H\subset W$ is
continuous.

{\it Proof.} It is clear that $H\subset W$ as a set. As $W$ is a
first-countable space, it is enough to check that if $\varphi _{n}\in H$ and 
$\varphi _{n}\longrightarrow 0$ in $H$, then $\varphi _{n}\longrightarrow 0$
in $W$. Now, $\varphi _{n}\longrightarrow 0$ in $H$ means $\varphi _{n}=\rho
(f_{n})$, where $f_{n}\in L_{2}^{\ast }(R)$ and $f_{n}\longrightarrow 0$ in 
$L_{2}^{\ast }(R)$. But then as $L_{2}^{\ast }(R)\subset W^{\ast }$ is a
continuous embedding, $f_{n}\longrightarrow 0$ in $W^{\ast }$. Also, we have
seen that $\rho (W^{\ast })=\widetilde{W}$ is a topological subspace of $W$.
Therefore $\varphi _{n}=\rho (f_{n})\longrightarrow 0$ in $W$.

{\it Theorem.} The embedding $W^{\ast }\subset H^{\ast }$ is continuous.

{\it Proof.} We assume that topology on $H^{\ast }$ is strong. It is
clear that $W^{\ast }\subset H^{\ast }$ as the set because any functional
continuous on $W$ is continuous on $H\subset W$. Any neighborhood $O$ of
zero in $H^{\ast }$ is a set of functionals $f\in H^{\ast }$ such that 
$\left\| f\right\| _{H^{\ast }}={\sup_{\psi \in B}}\left| (f,\psi
)\right| <{\epsilon} $, where $B$ is a unit ball in $H$. As the embedding 
$H\subset W$ is continuous, $B$ is bounded in $W$ as well. Consider the
neighborhood $\widetilde{O}$ in $W^{\ast }$ given by 
${\sup_{\psi \in A}}\left| (f,\psi )\right| <{\epsilon} $ with $A \in B$. Clearly then 
$\widetilde{O}\subset O$.

As a result, we have a chain of topological embeddings: 
\begin{equation}
\label{chain}
H\subset W\subset L_{2}(R)\subset L_{2}^{\ast }(R)\subset W^{\ast }\subset
H^{\ast }.
\end{equation}
Here we identify $L_{2}(R)$ and $L_{2}^{\ast }(R)$ by identifying in the
usual way each $L_{2}(R)$-function $\psi $ with the corresponding regular
functional $(\psi ,\cdot )_{L_{2}}$. 

The meaning of this chain will be discussed in the conclusion. 
For now let us use the fact that $H^{\ast}$ 
includes all functionals from $W^{\ast }$.
Consider the Hilbert space $H^{\ast}$ as a coordinate space
for the string space $\bf {S}$. Let $\Phi \in \bf {S}$ be a string
with coordinate function equal to the $\delta $- function. Since $H^{\ast }$
is Hilbert, the square of the $\delta $- function must be finite. Indeed, 
using (\ref{dualmetric}) with $\rho=\omega ^{-1}$ we have:

\begin{equation}
(\Phi ,\Phi )_{S}=(\delta ,\delta )_{H^{\ast }}=(\rho \rho ^{*}\delta
,\delta )=\int e^{-(x-y)^{2}-x^{2}}e^{-(y-z)^{2}-z^{2}}\delta (x)\delta
(z)dydxdz=\int e^{-4y^{2}}dy=\frac{\sqrt{\pi }}{2}.
\end{equation}
The fact that the norm of the $\delta $- function in $H^{\ast }$ is finite is of
course related to the fact that the metric
$g_{xz}=\int e^{-(x-y)^{2}-x^{2}}e^{-(y-z)^{2}-z^{2}}dy$ is a smooth
function of $x$ and $z$ and is capable of ``compensating'' singularities of
the product of two $\delta $- functions. In the case of $L_{2}$ spaces the
metric $g_{xz}$ is equal to $\delta (x-z)$. Therefore $\delta $-function can
not be a coordinate of a string and the norm of it is not defined.

Assume instead, that a coordinate space consists only of the smooth
functions. That is, it is obtained by some smoothing of the elements of 
$L_{2}$ (as the space $H$ in the example). Then the metric in the new
coordinates is obtained by a singularization of the $\delta (x-z)$ metric on 
$L_{2}$.

\section{Generalized eigenvalue problem} 
\setcounter{equation}{0}

Let $A$ be a linear operator on a linear topological space $V$. If $V$ is
finite-dimensional and $A$ is, say, Hermitian, then a basis in $V$ exists,
such that each vector of it is an eigenfunction of $A$. If $V$ is
infinite-dimensional, this statement is no longer true. Yet quite often
there exists a complete system of ``generalized eigenfunctions" of $A$ in the sense
of the definition below (see \cite{Gel-Kos}):

{\it Definition}.  A linear functional $F$ on $V$, such that 

\begin{equation}
F(A\Phi )=\lambda F(\Phi \bf{)}
\end{equation}
for every $\Phi \in V$, is called a {\it generalized eigenfunction of $A$
corresponding to the eigenvalue $\lambda $}. 

Assume now that $V$ is the
string space $\bf {S}$ and $e_{H}$ is a string basis on $\bf{
S}$. Assume $F$ is a generalized eigenfunction of a linear operator $\bf{
A}$ on $\bf {S}$. Then

\begin{equation}
F({\bf{A}}e_{H}\varphi )=\lambda F(e_{H}\varphi ),
\end{equation}
where $e_{H}\varphi =\Phi $. Therefore 

\begin{equation}
e_{H}^{\ast }F(e_{H}^{-1}{\bf{A}}e_{H}\varphi )=\lambda e_{H}^{\ast
}F(\varphi ).
\end{equation}
Here $e_{H}^{-1}{\bf{A}}e_{H}$ is the representation of $\bf{A}$ in
the basis $e_{H}$.

By defining $e_{H}^{\ast }F=f$ we have

\begin{equation}
\label{eigenvalue}
f(e_{H}^{-1}{\bf{A}}e_{H}\varphi )=\lambda f(\varphi ),  
\end{equation}
or:
\begin{equation}
\label{eigen}
f(A\varphi )=\lambda f(\varphi ),
\end{equation}
where $A=e_{H}^{-1}{\bf{A}}e_{H}$.

Notice that the last equation desribes not just one eigenvalue problem, but
a family of such problems, one for each string basis $e_{H}$. As we change 
$e_{H}$, the operator $A$ in general changes as
well, as do the eigenfunctions $f$.

{\it Example.} Let $W$ be the Schwartz space. Although it is not possible to
introduce a Hilbert metric on $W$, this will not play any role in what
follows. If necessary, one could replace $W$ with the Hilbert space $H\subset W$ 
from the previous example. Consider the operator of
differentiation $A=i\frac{d}{dx}$ on $W$. The generalized eigenvalue problem
for $A$ is 

\begin{equation}
\label{eigenvalue1}
f(i\frac{d}{dx}\varphi )=\lambda f(\varphi ),  
\end{equation}
where $f\in W^{\ast }$. \ The functionals 
\begin{equation}
f(x)=e^{-i\lambda x}
\end{equation}
are the eigenvectors of $A$. Let us now consider a coordinate change $\rho
:W\longrightarrow W$ given by the Fourier transform. That is, 

\begin{equation}
\psi (k)=(\rho \varphi )(k)=\int \varphi (x)e^{ikx}dx.
\end{equation}
The Fourier transform is a toplinear automorphism of $W$. The inverse transform
is given by

\begin{equation}
(\omega \psi )(x)=\frac{1}{2\pi }\int \psi (k)e^{-ikx}dk.
\end{equation}
According to (\ref{eigenvalue}) the eigenvalue problem in new coordinates is

\begin{equation}
\omega ^{\ast }f(\rho A\omega \psi )=\lambda \omega ^{\ast }f(\psi ).
\end{equation}
We have

\begin{equation}
A\omega \psi =i\frac{d}{dx}\frac{1}{2\pi }\int \psi (k)e^{-ikx}dk=\frac{1}{
2\pi }\int k\psi (k)e^{-ikx}dk.
\end{equation}
Therefore,

\begin{equation}
(\rho A\omega \psi )(k)=k\psi (k).
\end{equation}
So, the eigenvalue problem in new coordinates is as follows:

\begin{equation}
\label{eigenvalue2}
g(k\psi )=\lambda g(\psi ).  
\end{equation}
Thus, we have the eigenvalue problem for the operator of multiplication by
the variable. The eigenvectors here are given by 
\begin{equation}
g(k)=\delta (k-\lambda ).
\end{equation}
Notice that $g=\omega ^{\ast }f$ is as it should be. Indeed, 
\begin{equation}
(\omega ^{\ast }f)(k)=\frac{1}{2\pi }\int f(x)e^{ikx}dx=\frac{1}{2\pi }\int
e^{-i\lambda x}e^{ikx}dx=\delta (k-\lambda ).
\end{equation}

As a result, the eigenvalue problems (\ref{eigenvalue1}), and (\ref
{eigenvalue2}) are two coordinate expressions of a single eigenvalue problem

\begin{equation}
F({\bf{A}}\Phi )=\lambda F(\Phi) 
\end{equation}
for an operator $\bf{A}$ on $\bf {S}$.

If $H$ is a Hilbert space, then any functional $f$ on $H$ is given by 
\begin{equation}
f(\varphi)=(\psi,\varphi)_{H}.
\end{equation}
Therefore, the generalized eigenvalue problem (\ref{eigen}) for an operator
$A$ on $H$ can also be written in the form
\begin{equation}
\label{evg}
(\psi, A\varphi)_{H}=\lambda(\psi,\varphi)_{H}. 
\end{equation}
The last equation must be true for any function $\varphi \in H$.

If $A^{+}$ is the Hermitian conjugate of $A$, then (\ref{evg}) gives
\begin{equation}
\label{cross}
(A^{+}\psi,\varphi)_{H}=\lambda(\psi,\varphi)_{H}
\end{equation}
for any $\varphi \in H$. Assume now that $H$ is a space of ordinary (i.e. not generalized)
functions. Then from (\ref{cross}) we have
\begin{equation}
A^{+}\psi=\lambda\psi.
\end{equation}
That is, $A$ has an eigenvalue $\lambda$ in generalized sense if and only if 
$A^{+}$ has $\lambda$ as an eigenvalue in the ordinary sense.

If $(\psi,\varphi)_{H}=({\widehat{G}}\psi,\varphi)$, then 
\begin{equation}
({\widehat{G}}\psi,A\varphi)=(A^{*}{\widehat{G}}\psi,\varphi)=({\widehat{G}}A^{+}\psi,\varphi).
\end{equation}
This yields the following relationship between the operators:
\begin{equation}
A^{+}={\widehat{G}}^{-1}A^{*}{\widehat{G}}.
\end{equation}

Having introduced generalized eigenvectors it is natural to ask whether we can make
a basis out of them. For this we introduce the following

{\it Definition}. A string basis $e_{H}$ is a {\it basis of eigenvectors} of 
a linear operator ${\bf{A}}:{\bf {S}}\longrightarrow {\bf {S}}$ with eigenvalues
$\lambda=\lambda(k)$ if 
\begin{equation}
\label{eb0}
{\bf {A}}e_{H}(\varphi)=e_{H}(\lambda\varphi)
\end{equation}
for any $\varphi \in H$.

Notice that in agreement with the general approach advocated here the basis
of eigenvectors is a linear map from $H$ onto $\bf {S}$ and $\lambda$ is a
function of $k$.

If $e_{H}$ is a basis of eigenvectors of $\bf {A}$, then
\begin{equation}
\label{eb}
(\Phi,{\bf {A}}\Psi)_{S}=(e_{H}\varphi,e_{H}(\lambda\psi))_{S}=
(\varphi,\lambda\psi)_{H}.
\end{equation}
In particular,
\begin{equation}
\label{eb1}
(\Phi,{\bf {A}}\Phi)_{S}=(\varphi,\lambda\varphi)_{H}.
\end{equation}
If $H=l_{2}$, this reduces to 
$\sum_{k}\lambda _{k}\left| \varphi _{k}\right| ^{2}$. 
If $H=L_{2}(R)$ instead, (\ref{eb1}) yields 
$\int \lambda (k)\left| \varphi (k)\right| ^{2}dk$.

By rewriting (\ref{eb0}) as
\begin{equation}
{e_{H}}^{-1}{\bf {A}}e_{H}(\varphi)=\lambda\varphi
\end{equation}
we see that the problem of finding a basis of eigenvectors of $\bf {A}$ is equivalent to 
the problem of finding such a string basis $e_{H}$ in which the action of $\bf {A}$
reduces to multiplication by a function $\lambda$. In particular case of an $l_{2}$-basis
this yields the classical problem of finding a basis of eigenvectors of a 
linear operator.

Notice that in the invariant approach advocated here we are not free to define $A$ on a
specific Hilbert space $H$. In fact, the choice of $e_{H}$ dictates not just 
$H$, but the specific coordinate representation $A$ of $\bf{A}$. On the
other hand, if the space $H$ and an operator $A$ on it are given, then by
changing the basis we can find the unique representation of $\bf{A}$ in
any basis.

The entire discussion seems to be very similar to the usual change of matrix representation
of a linear operator acting on a finite-dimensional vector space. There is,
however, an important difference. As in the finite-dimensional case a linear
operator acting on a Hilbert space $H$ can be represented by an (infinite)
matrix. For this we simply choose a basis on $H$ and find the matrix
representation of $A$ in this basis. What we are doing here is different.
Whenever $H$ is the space of functions, we consider these functions as
coordinates of invariant elements of the string space $\bf {S}$. This
passage to $\bf {S}$ permits us to consider all possible functional Hilbert
spaces at once as images of different coordinate charts on $\bf {S}$.
This is very useful for two reasons. First of all, it relates objects of
different nature by asserting that they are coordinate representations of
one and the same invariant object on the string space. Second, a particular
choice of a coordinate chart can simplify significantly the problem in hand.
We also see that the particular properties of functions in $H$ are no
longer important as we change these properties by changing coordinates.

\section{Nonlinear coordinate transformations on $\bf {S}$}

\setcounter{equation}{0}
Up until now we considered coordinate transformations that were isomorphisms
of coordinate Hilbert spaces. To have a differentiable manifold structure on 
$\bf {S}$ we need to consider local nonlinear diffeomorphisms of Hilbert
spaces as well.

{\it Theorem. }Let $\rho $ be a bijection of a Hilbert space $H$ onto a
Hilbert space $V=\rho (H).$ Then $\rho $ induces on $V$ a new metric
structure such that $V$ with this structure is a complete linear metric
space.

{\it Proof. }Given $f=\rho (\varphi )$ and $g=\rho (\psi )$ we define
the distance from $f$ to $g$ by

\begin{equation}
\label{metric}
d_{V}(f,g)=\left\| \varphi -\psi \right\| _{H}.  
\end{equation}
We clearly have 

\begin{eqnarray}
d_{V}(f,f) &=&0,\  d_{V}(f,g)>0\ for f\neq g\\
d_{V}(f,g) &=&d_{V}(g,f) \\
d_{V}(f,h) &\leq &d_{V}(f,g)+d_{V}(g,h).
\end{eqnarray}
That is, $d_{V}$ is indeed a metric. When the topology on $V$ is defined by 
$d_{V}$, the operator $\rho $ becomes a diffeomorphism from $H$ onto $V$
and V becomes a complete metric space.

In general we shall consider local diffeomorphisms of Hilbert spaces. That
is, $\rho :N\longrightarrow \widetilde{H}$ will be defined on a neighborhood 
$N$ of a point $\varphi \in H$. In this case using (\ref{metric}) and taking 
$h=f-g$ and $\omega =\rho ^{-1}$ we have: 

\begin{equation}
d_{V}^{2}(f,g)=\left\| \omega f-\omega g\right\| _{H}^{2}=(\widehat{G}
_{H}(\omega f-\omega g),\omega f-\omega g)=({\widehat{G}}_{H}(Lh+\alpha
h),Lh+\alpha h)=(L^{\ast }G_{H}Lh,h)+\beta h.
\end{equation}
Here ${\widehat{G}}_{H}:H\longrightarrow H^{\ast }$ defines the metric on $H$,
the linear isomorpism $L:\widetilde{H}\longrightarrow H$ is the derivative
of $\omega $, and $\alpha ,\beta \longrightarrow 0$ as $\left\| h\right\| _{
\widetilde{H}}\longrightarrow 0$. Thus, given the string bases $e_{H}$ and 
$e_{\widetilde{H}}$, the operator

\begin{equation}
\label{newmetric}
\widehat{G}_{\widetilde{H}}=L^{\ast }{\widehat{G}}_{H}L  
\end{equation}
defines a Hilbert metric on the tangent space $T_{g}\widetilde{H}$ to 
$\widetilde{H}$ at the point $g.$

\section{String manifolds}
\setcounter{equation}{0}

We define a string manifold $\it {S}$\ as a manifold modelled on $\bf{
S}$ and furnished with a coordinate structure (see below) and a Riemannian
metric. Before introducing a notion of coordiante structure let us recall
the definition of a differential manifold that works in infinte-dimensional
case. See \cite{Lang} for details.

{\it Definition}. Let $\it {S}$ be a set. Let $U_{\alpha }$ (with $
\alpha $ changing in some indexing set) be a collection of subsets of $
\it {S}$ that covers $\it {S}.$ For each $\alpha $ let $\pi _{\alpha }$
be a bijection of $U_{\alpha }$ onto an open subset $\pi _{\alpha
}(U_{\alpha })$ of $\bf {S}.$ Assume that for any $\alpha ,\beta ,\pi
_{\alpha }(U_{\alpha }\cap U_{\beta })$ is open in $\bf {S}$. Assume also
that for each pair of indices $\alpha ,\beta $ the map $\pi _{\beta }\pi
_{\alpha }^{-1}:\pi _{\alpha }(U_{\alpha }\cap U_{\beta })\longrightarrow
\pi _{\beta }(U_{\alpha }\cap U_{\beta })$ is a $C^{\infty }$-isomorphism.
Then the collection of pairs $(U_{\alpha },\pi _{\alpha })$ is called a {\it
$C^{\infty }$ -atlas} on $\it {S}$.

Notice that by taking $U_{\alpha }$ to be open, we give $\it {S}$ a topology. In
this topology $\pi _{\alpha }$ are homeomorphisms.

{\it Definition}.  Each pair $(U_{\alpha },\pi _{\alpha })$ is called a
{\it chart} of the atlas.

{\it Definition}.  Given an open set $U\subset \it {S}$ and a
homeomorphism $\pi :U\longrightarrow \bf{U}$ onto an open set of 
$\bf {S}$ we say that $(U,\pi )$ is {\it compatible} with the atlas $(U_{\alpha
},\pi _{\alpha })$, if each map $\pi _{\alpha }\pi ^{-1}$, whenever defined,
is a diffeomorphism.

{\it Definition}.  Two atlaces are {\it compatible} if each chart of one is
compatible with the other atlas.

Compatibility is an equivalence relation between the atlaces.

{\it Definition}. An equivalence class of $C^{\infty }$-atlaces with
respect to compatibility relation is said to define a structure of an
abstract $C^{\infty }$-{\it Hilbert} manifold.

So locally $\it {S}$ looks like $\bf {S}$. Let us now introduce a
coordinate formalism on $\it {S}$.

{\it Definition}. Let $(U_{\alpha },\pi _{\alpha })$ be an atlas on 
$\it {S}.$ Consider a collection of quadruples $(U_{\alpha },\pi _{\alpha
},\omega _{\alpha },H_{\alpha })$, where each $H_{\alpha }$ is a Hilbert
space of functions and $\omega _{\alpha }$ is an isomorphism of $\bf {S}$
onto $H_{\alpha }$. Such a collection will be called a {\it functional atlas} on 
$\it {S}$. A collection of all compatible functional atlaces on $\it {S}
$ will be called a {\it coordinate structure} on $\it {S}.$

That is, $\it {S}$ with a coordinate structure can be thought of as a
Hilbert manifold $\it {S}$\ with differentiable structure defined by the
atlaces $(U_{\alpha },\omega _{\alpha }\circ \pi _{\alpha })$. We prefer,
however, to distinguish between an abstract Hilbert manifold and \ a Hilbert
manifold with a coordinate structure.

Let $(U_{\alpha },\pi _{\alpha })$ be a chart on $\it {S}$.
For each $p\in U_{\alpha }$, $\pi _{\alpha }(p)\in \bf {S}$. Usually
one introduces the $i$-th\ coordinate function $p^{i}$ on a Hilbert manifold
by choosing a basis $\{e_{i}\}$\ on $\bf {S}$ and taking the $i$-th
component of $\pi _{\alpha }(p)$ in the basis. Thus, we obtain coordinate
functions by identifying $\bf {S}$ with the Hilbert space $l_{2}$ of
sequences. The coordinate structure in the definition above is more general
in that we are able to identify $\bf {S}$ with any Hilbert space of
functions and not only $l_{2}$. In this case, if $p\in U_{\alpha },$ then 
$\omega _{\alpha }\circ \pi _{\alpha }(p)$ will be called the {\it coordinate map}
or simply the {\it coordinate} of $p$. The isomorphisms 
$\omega _{\beta }\circ \pi _{\beta }\circ (\omega _{\alpha }\circ 
\pi _{\alpha })^{-1}:\omega _{\alpha }\circ \pi _{\alpha }
(U_{\alpha }\cap U_{\beta })\longrightarrow 
\omega _{\beta }\circ \pi _{\beta }(U_{\alpha }\cap U_{\beta })$ are called
{\it coordinate transformations} on $\it S$.

As $\it {S}$ is a differentiable manifold we can introduce the tangent
bundle structure $\tau :T{\it{S}}\longrightarrow {\it {S}}$ and the bundle $\tau
_{s}^{r}:T_{s}^{r}{\it {S}}\longrightarrow {\it {S}}$
of tensors of rank $(r,s)$. Consider in particular the bundle 
$\tau _{2}:ST_{2}\it {S}\longrightarrow \it {S}$ of symmetric $(0,2)$
tensors.

{\it Definition}. A {\it Riemannian metric} on $\it {S}$ is a section $g:
{\it {S}}\longrightarrow ST_{2}{\it {S}}$ of $\tau _{2}:ST_{2}{\it {S}}
\longrightarrow \it {S}$, such that $g_{p}$ is positive definite for
every $p\in \it {S}$, i.e. $g_{p}(\Phi ,\Phi )>0$ for every $p\in {\it{
S}}$ and any $\Phi \in T_{p}\it {S}$. Here $T_{p}\it {S}$ is the
tangent space to $\it {S}$ at $p.$

{\it Definition}.  A Hilbert manifold with coordinate structure and
Riemannian metric is called a {\it string manifold}.

Coordinate structure on a string manifold permits one to obtain a functional
description of any tensor. Namely, let ${\bf{G}}_{p}(F_{1},...,F_{r},\Phi
_{1},...,\Phi _{s})$ be an $(r,s)$-tensor on $\it {S}$.

{\it Definition}. The coordinate map $\omega _{\alpha }\circ \pi _{\alpha
}:U_{\alpha }\longrightarrow H_{\alpha }$ for each $p\in U_{\alpha }$\ yelds
the linear map of tangent spaces $d\rho _{\alpha }:T_{\omega _{\alpha }\circ
\pi _{\alpha }(p)}H_{\alpha }\longrightarrow T_{p}\it {S}$, where $\rho
_{\alpha }=\pi _{\alpha }^{-1}\circ \omega _{\alpha }^{-1}.$ This map is
called a {\it local coordinate string basis on} $\it {S}$.

Let $e_{H_{\alpha }}=e_{H_{\alpha }}(p)$ be such a basis and $e_{H_{\alpha
}^{\ast }}=e_{H_{\alpha }^{\ast }}(p)$ be the corresponding dual basis.
Notice that for each $p$ the map\ $e_{H_{\alpha }}$ is a string basis as
defined in section 2. Therefore, the local dual basis is defined for each $p$
as before and is a function of $p.$

We now have $F_{i}=e_{H_{\alpha }^{\ast }}f_{i}$, and $\Phi
_{j}=e_{H_{\alpha }}\varphi _{j}$ for any $F_{i}\in T_{p}^{\ast }\it {S}$
, $\Phi _{j}\in T_{p}\it {S}$ and some $f_{i}\in H_{\alpha }^{\ast
},\varphi _{j}\in H_{\alpha }$. Therefore ${\bf{G}}_{p}(F_{1},...,F_{r},
\Phi _{1},...,\Phi _{s})=G_{p}(f_{1},...,f_{r},\varphi _{1},...,\varphi _{s})
$ defining component functions of the $(r,s)$-tensor ${\bf{G}}_{p}$ in the
local coordinate basis $e_{H_{\alpha }}$.

\section{Concluding remarks}

\setcounter{equation}{0}

The notion of an infinite-dimensional manifold is a direct generalization
of its finite dimensional counterpart. Yet many techniques available to us in 
a finite dimensional setting can not be easily generalized to the case of infinite
dimensions.

Perhaps this is so because we are trying to generalize in a wrong way. To explain,
consider for example a passage from the space of $n$-columns $R^{n}$ to the Hilbert
space of sequences $l_{2}$. A significant difference between these two spaces is 
the notion of convergence needed to identify a sequence of numbers as an element of 
$l_{2}$. This notion is exactly the new entity related to infinite dimensionality of $l_{2}$.
In setting up a differentiable structure we do not pay that much attention to the
kind of convergence available on the model space $l_{2}$. Even when, say,
several Hilbert spaces of sequences are simultaneously considered as models of a
Hilbert manifold, differentiable structure is not affected by the difference
between them. It is concerned only with the differentiability of maps.

It is advocated here that a much more productive approach to infinite-dimensional
manifolds is to use the difference between Hilbert models even
if the manifold structure itself is not altered by changing a model.

A particular choice of a model is significant in applications where properties of functional
objects depend on the type of space of functions used. By considering several models at
once we can reduce the seemingly unrelated problems on different spaces of functions 
to equivalence classes of problems on the string space. The generalized eigenvalue 
problem considered in section 5 provides an example.

We also obtain the possibility to
reformulate a problem given on one space in terms of another space. A good example
of the usefulness of such a reformulation is provided by the theory of generalized
functions (distributions). In this theory operations that could not be defined
by themselves on spaces containing singular distributions are defined first on fundamental 
spaces of ``good" functions. Then they are ``transplanted" to the much
larger dual spaces. 

In the approach advocated here this passage from a space to its 
dual is a coordinate transformation and the operations themselves could be defined in 
invariant manner on the string space. 

Notice also, that a coordinate transformation on $\bf S$ can alter analytic
properties of elements of the coordinate space by changing, for example, singular distributions
to infinite differentiable functions and vice versa. This was explored in example of 
section 4.

Choosing appropriate coordinates on $\bf S$ for a problem in hand is as useful (if
not more) as
choosing canonical coordinates for a finite dimensional problem. For example, Fourier and Laplace
transforms reduce differentiation of functions to multiplication by a variable 
providing an algebraic approach to solving differential equations.

We also see that in this approach the finite and infinite-dimensional manifolds become
related in a new way. The notions of a string basis, dual string basis,
orthogonal string basis, local coordinate basis are the clear analogues of their
finite dimensional counterparts. Simultaneously they provide us with the power of changing
the coordinate spaces and the corresponding functional description of invariant
objects (tensors).

The main spaces of functions used in the paper are Hilbert spaces. The role of countably
normed spaces like the Schwartz space $W$ in this setting is not completely understood.
The symmetric way in which $H$ and $W$ appear in the chain (\ref{chain})
makes us think that such spaces should be considered as possible coordinate spaces as
well. This is, however, the subject for a different paper.

\end{document}